\documentclass[10pt,conference]{IEEEtran}
\IEEEoverridecommandlockouts
\usepackage{cite}
\usepackage{amsmath,amssymb,amsfonts}
\usepackage{algorithmic}
\usepackage{algorithm}
\usepackage{graphicx}
\usepackage{varwidth}
\usepackage{textcomp}
\usepackage{xcolor}
\usepackage{wrapfig}
\usepackage{pifont}
\usepackage{booktabs}
\usepackage{colortbl}
\usepackage{soul}
\usepackage[hyphens]{url}
\usepackage{hyperref}
\usepackage{multirow}

\def\BibTeX{{\rm B\kern-.05em{\sc i\kern-.025em b}\kern-.08em
    T\kern-.1667em\lower.7ex\hbox{E}\kern-.125emX}}
\begin{document}

\title{QuantumSEA: In-Time \underline{S}parse \underline{E}xploration for Noise \underline{A}daptive Quantum Circuits}
\author{\normalsize{QCE 2023 Submission
    \textbf{\#121} -- Confidential Draft -- Do NOT Distribute!!}}


\author{
Tianlong Chen\textsuperscript{1}, 
Zhenyu Zhang\textsuperscript{1},
Hanrui Wang\textsuperscript{2},
Jiaqi Gu\textsuperscript{1,5},
Zirui Li\textsuperscript{3}, \\
David Z. Pan\textsuperscript{1},
Frederic T. Chong\textsuperscript{4},
Song Han\textsuperscript{2},
Zhangyang Wang\textsuperscript{1} \\
{\textsuperscript{1}University of Texas at Austin} 
{\textsuperscript{2}Massachusetts Institute of Technology} \\
{\textsuperscript{3}Rutgers University}
{\textsuperscript{4}University of Chicago}
{\textsuperscript{5}Arizona State University}
}

\maketitle

\begin{abstract}
Parameterized Quantum Circuits (PQC) have obtained increasing popularity thanks to their great potential for near-term Noisy Intermediate-Scale Quantum (NISQ) computers. Achieving quantum advantages usually requires a large number of qubits and quantum circuits with enough capacity. However, limited coherence time and massive quantum noises severely constrain the size of quantum circuits that can be executed reliably on real machines. To address these two pain points, we propose QuantumSEA, an in-time \underline{s}parse \underline{e}xploration for noise-\underline{a}daptive quantum circuits, aiming to achieve two key objectives: ($1$) \textit{implicit circuits capacity during training} - by dynamically exploring the circuit's sparse connectivity and sticking a fixed \emph{small} number of quantum gates throughout the training which satisfies the coherence time and enjoy light noises, enabling feasible executions on real quantum devices; ($2$) \textit{noise robustness} - by jointly optimizing the topology and parameters of quantum circuits under real device noise models. In each update step of sparsity, we leverage the moving average of historical gradients to grow necessary gates and utilize salience-based pruning to eliminate insignificant gates. Extensive experiments are conducted with $7$ Quantum Machine Learning (QML) and Variational Quantum Eigensolver (VQE) benchmarks on $6$ simulated or real quantum computers, where QuantumSEA consistently surpasses noise-aware search, human-designed, and randomly generated quantum circuit baselines by a clear performance margin. For example, even in the most challenging on-chip training regime, our method establishes state-of-the-art results with only \textbf{half} the number of quantum gates and $\boldsymbol{\sim2\times}$ time saving of circuit executions. Codes are available at \url{https://github.com/VITA-Group/QuantumSEA}.
\end{abstract}


\section{Introduction}
As a rising computational paradigm, Quantum computing (QC) demonstrates great potential in solving traditionally intractable problems with exponential acceleration over classical computers. Numerous domains have witnessed rapid progress of QC, including cryptography~\cite{shor1999polynomial}, database search~\cite{grover1996fast}, chemistry~\cite{cao2019quantum,kandala2017hardware,peruzzo2014variational}, and machine learning~\cite{biamonte2017quantum,farhi2014quantum,harrow2009quantum,lloyd2013quantum,rebentrost2014quantum}. Among these, parameterized Quantum Circuits (PQCs)~\cite{beer2020training} on the noisy intermediate-scale quantum (NISQ) devices~\cite{bharti2021noisy} is one of the most promising near-term applications showing impressive benefits from QC.

\begin{figure}[t]
    \centering
    \includegraphics[width=1\linewidth]{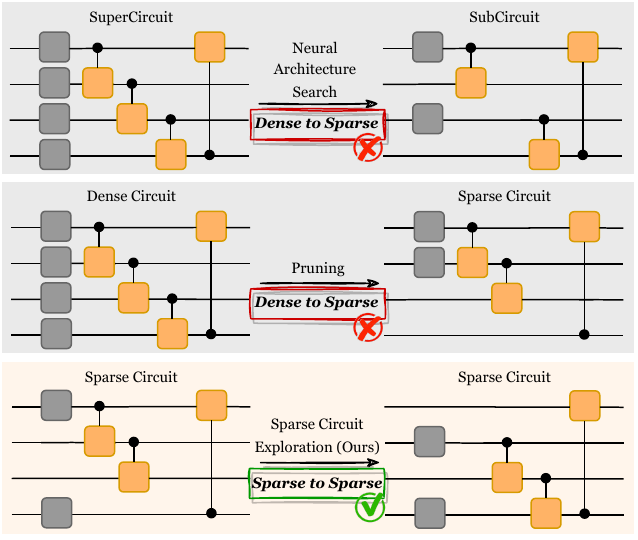}
    \vspace{-2mm}
    \caption{Overview of method comparisons. Both neural architecture search and pruning maintain dense circuits during training and produce their sparse variants in the end. And their training suffers from massive decoherence and operation noises due to excess quantum gates. Our proposal enables the circuit design from sparse to sparse, which only deals with a small portion of quantum gates at each training iteration.}
    \label{fig:teaser_methods}
    \vspace{-2mm}
\end{figure}

To reach PQC's quantum advantage, it demands a large number of qubits and quantum circuits with sufficient capacity~\cite{wang2022chip}. Most of the existing works~\cite{harrow2009quantum,magesan2012characterizing} that perform both training and inference on classical computers via software simulations, can not satisfy these conditions, since they have poor scalability and are failed to deal with exponentially boosted time and memory costs ($\mathcal{O}(2^{\#\mathrm{qubits}})$)~\cite{wang2022chip}. Unfortunately, two challenging impediments arise when we consider the on-chip training or evaluation on real quantum machines. Firstly, the unwanted interactions between qubits and the environment or other qubits cause qubit decoherence. The \textit{coherence time} statistically indicates how long a qubit retains its information, i.e., a qubit's ``lifetime''. Therefore, all operations are required to be completed before significant decoherence happens. Otherwise, the decoherence will cause severe infidelity of quantum-information processing~\cite{wang2021single}, and stultify the computing results. Therefore, the \textit{depth} of PQC should be small enough to alleviate the decoherence impact. 
Meanwhile, the non-negligible \textit{quantum errors} in the current NISQ era also make operations on qubits highly unreliable. For instance, the quantum gates commonly suffer from a high error rate of $10^{-3}$ to $10^{-2}$~\cite{wang2021quantumnas, wang2022chip}. In the context of PQC on real QC devices, such noises will result in unstable training and also significantly degraded performance~\cite{wang2022chip} compared to the ideal noise-free simulation. Thus, to bridge quantum algorithms to realistic QC hardware, \textit{robust circuits topology} against quantum noises are on the urgent call.

\begin{figure*}[t]
    \centering
    \includegraphics[width=0.95\linewidth]{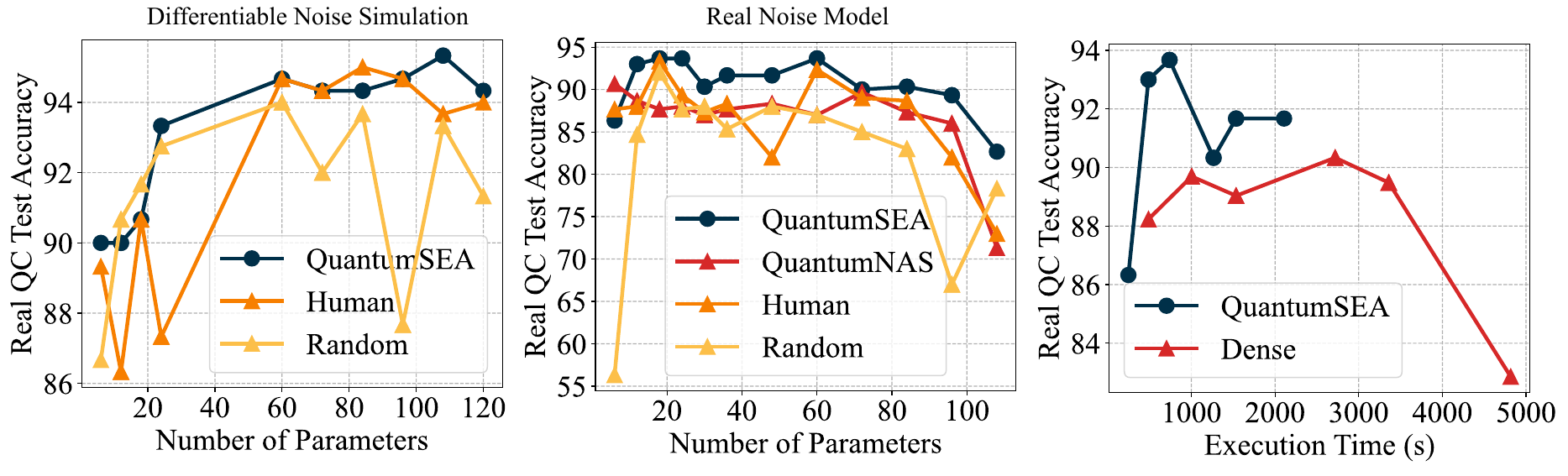}
    \vspace{-5mm}
    \caption{\textit{(Left and Middle)} Real QC testing accuracy (\%) over the number of parameters in quantum circuits. Evaluations are conducted on \texttt{IBMQ-Santiago} with \texttt{RXYZ} and MNIST-2. The accuracy of conventionally designed circuits is quickly saturated and then degraded. Our QuantumSEA outperforms other approaches in quantum noise mitigation, allowing larger circuit capacity and superior accuracy. (\textit{Right}) Real QC testing accuracy (\%) over the training time of quantum circuits, \textit{i.e.}, \texttt{RXYZ}) on \texttt{IBMQ-Santiago}.} 
    \label{fig:acc_para_noise_teaser}
    \vspace{-2mm}
\end{figure*}

To address these obstacles, we propose QuantumSEA, an in-time \textbf{S}parse \textbf{E}xploration for noise-\textbf{A}daptive robust quantum circuits. \underline{On one hand}, our proposal starts with a randomly sparsified circuit. It dynamically explores its connectivity across the training time, while sticking to a fixed small number of quantum gates. The reduced quantum operations effectively decrease noises and alleviate the negative effects of decoherence. It is also one of the key benefits of our proposal compared to the conventional circuit designs from neural architecture search or pruning, as demonstrated in Figure~\ref{fig:teaser_methods}. Meantime, the dynamic topology exploration brings extra implicit circuit capacity in the space-time manifold, achieving significant improvements in the expressibility of sparse quantum circuits. \underline{On the other hand}, QuantumSEA develops the robust circuit by \textit{simultaneously} optimizing its sparse connectivities and parameters under real QC noises. Moreover, identified topologies are capable of serving an implicit regularization and promoting noise resistance, as widely demonstrated in classical machine learning~\cite{ye2019adversarial,gui2019model,ding2022audio}. In other words, QuantumSEA kills two ``birds'' (short coherent time and large quantum noises) with one ``stone'' (sparsity), i.e., the located noise-adaptive sparse circuit topology enables robust \& efficient inference and training of PQCs on real QC machines with substantially improved performance and reduced execution time, as presented in Figure~\ref{fig:acc_para_noise_teaser} (\textit{Right}). Our innovations are unfolded along with three thrusts below:

\begin{itemize}
    \item [$\star$] \textbf{In-time sparse explorations} grant the quantum circuit extra implicit capacity, which significantly improves its expressiveness without introducing additional gates. Compared to the dense counterpart, our proposal is capable of reducing \textbf{over $2\times$} execution time (Figure~\ref{fig:acc_para_noise_teaser} (\textit{Right})), lowing the risk of severe decoherence, and improving the fidelity of predictions. \vspace{1mm}
    \item [$\star$] \textbf{Noise adaptive and jointly optimized topology \& weights} lead to a superior noise-resilient quantum circuit, outperforming noise-aware searched, random generated, and human-designed quantum circuit variants, as demonstrated in Figure~\ref{fig:acc_para_noise_teaser} (\textit{Left and Middle}). \vspace{1mm}
    \item [$\star$] \textbf{Extensive experiments conducted on real quantum devices} consistently demonstrate the effectiveness of our proposal. In total, $7$ representative benchmarks in QML and VQE on $6$ real quantum machines are evaluated, where we establish a new start-of-the-art on-chip training performance with significant running time savings based on the recent advance~\cite{wang2022chip}. Take MNIST-2 as an example. We achieve $1.33\%\sim 5.32\%$ accuracy gains and $\sim 50\%$ execution time reduction.
\end{itemize}

\section{Related Works}

\paragraph{Quantum machine learning (QML) and variational quantum eigensolver (VQE)} 

QML~\cite{biamonte2017quantum,lloyd2016quantum,lloyd2014quantum,lloyd2020quantum,mitarai2018quantum,cincio2021machine}studies machine learning algorithms on quantum computers, which shows an impressive ability in generating and estimating highly complex kernels~\cite{havlivcek2019supervised}. Recently, a group of QML models, QNN~\cite{beer2020training}, emerged, which parameterizes quantum circuits with trainable weights. Although diverse theoretical formulations of QNNs have been established, most of their implementations are limited to small-scale simulations of quantum systems via classical computing~\cite{farhi2018classification}. VQE~\cite{peruzzo2014variational,mcclean2016theory,tilly2021variational} is another representative example of NISQ algorithms, whose goal is to estimate the Hamiltonian's ground-state energy, serving as the first step towards investigating the energetic properties of molecules~\cite{deglmann2015application,williams2018free,heifetz2020quantum} and materials~\cite{continentino2021key,van2020rechargeable}. VQE leverages the superposition principle~\cite{silverman2008quantum} of quantum mechanics to encode exponentially increased information with only a linearly grown number of qubits.

\paragraph{Quantum circuit connectivity design with QML} Both~\cite{he2022search} and~\cite{wang2021quantumnas} adopt neural architecture search approaches to automate the connectivity design of quantum circuits. The former~\cite{he2022search} introduces a space pruning algorithm to trim down the search space during architecture exploration by progressively removing unpromising candidate gates; the later~\cite{wang2021quantumnas} further uses magnitude-based pruning methods to eliminate insignificant quantum gates of searched quantum circuits. More possibilities for the circuit design via advanced search algorithms have been extensively discussed~\cite{zhang2020differentiable,ostaszewski2021reinforcement,kuo2021quantum,du2022quantum,bilkis2021semi}. It is worth mentioning that our QuantumSEA is fundamentally different from neural architecture based approaches. Since we always optimize a highly reduced number of quantum gates, QuantumSEA (1) avoids the massive operation noises from additional quantum gates (2) and makes quantum circuits more feasible for on-chip training given the limitation of qubit's coherence time.


To tackle the challenging optimization in the vast parameter space of large quantum circuits, \cite{sim2021adaptive} proposes a parameter-efficient circuit training framework by leveraging the adaptive pruning techniques from classical machine learning. In each iteration, it actives and optimizes a selected subset of circuit parameters, which dynamically alters the circuit connectivity. But they adopt a random strategy to update the circuit structure, resulting in a sub-optimal optimization process.\cite{ostaszewski2021structure} also evolves the circuit structure together with the optimization of circuit parameters by simultaneously selecting the gate and finding the direction \& angle of rotation that yield minimum object values, in which they maintain all the gates during training, under the risk of more gate errors. Moreover, these methods only demonstrate the effectiveness of their proposal in small-scale cases like parameterized single-qubit gates and fixed two-qubit gates.

\paragraph{Quantum noise and its mitigation} Quantum noises are widely existing in real quantum computer devices, resulting from unwanted interactions between qubits, imperfect control signals, or interference from the environment~\cite{hsieh2020realistic}. And quantum noise mitigation has been extensively developed recently. Extrapolation methods~\cite{temme2017error,li2017efficient} extrapolate the ideal noise-free measurement from multiple measurements of a quantum circuit with different noise rates. Quasiprobability~\cite{temme2017error,huo2021self,wang2021roqnn} inserts \texttt{X}, \texttt{Y}, and \texttt{Z} quantum gates to the circuit in probability and then calculates the summation to cancel out the noise effects. Quantum subspace expansion~\cite{mcclean2017hybrid}, learning-based mitigation~\cite{strikis2021learning,czarnik2021error, wang2022quest}, post-measurement normalization and quantization~\cite{wang2021roqnn} are invented to address the noise issue. In our work, we proposed QuantumSEA to neutralize noise impacts from a fundamentally different perspective: high-quality sparse typologies of quantum circuits. 


\definecolor{lemonchiffon}{rgb}{1.0, 0.98, 0.8}
\definecolor{babyblueeyes}{rgb}{0.63, 0.79, 0.95}

\begin{figure*}[t]
    \centering
    \includegraphics[width=0.9\linewidth]{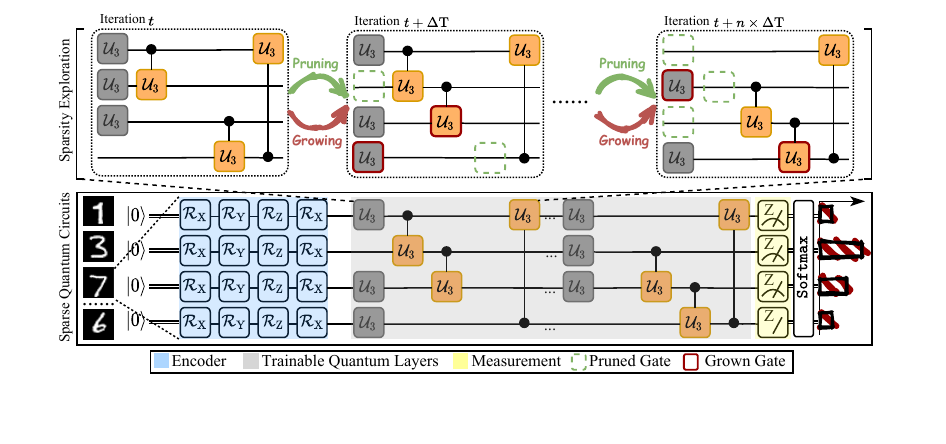}
    \vspace{-4mm}
    \caption{The overview of QuantumSEA. The \textit{upper} figure demonstrates the in-time sparse exploration of quantum circuits. Specifically, it first trains sparse circuit for $\Delta\mathrm{T}$ iterations, then leverages prune-and-grow strategies to explore the crucial sparse topology, repeating until convergence. Both weight and topology updates are aware of real QC noises. The \textit{bottom} figure presents an example sparse circuit for QML tasks, which consists of \colorbox{babyblueeyes}{data encoder}, \colorbox{lightgray}{trainable quantum layers}, and \colorbox{lemonchiffon}{the measurement layer}.} 
    \label{fig:framework}
    \vspace{-5mm}
\end{figure*}

\begin{figure}[t]
    \centering
    \includegraphics[width=1\linewidth]{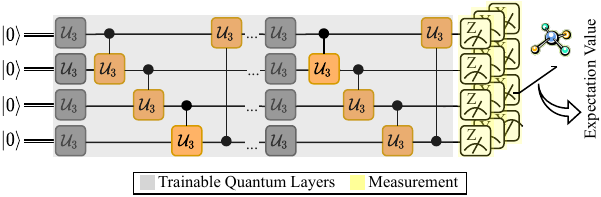}
    \vspace{-5mm}
    \caption{\small Example quantum circuits for VQE tasks.} 
    \vspace{-6mm}
    \label{fig:vqe_qnn}
\end{figure}

\section{Methodology}

\subsection{Preliminary}

\paragraph{Qubits} A quantum bit $|\psi\rangle$ is defined as a linear combination of two basis states $|0\rangle$ and $|1\rangle$: $|\psi\rangle=\alpha \times |0\rangle + \beta \times |1\rangle$, where $\alpha,\beta\in\mathbb{C}$ and $|\alpha|^2+|\beta|^2=1$. Leveraging the superposition principles~\cite{silverman2008quantum}, an $n$-qubits quantum system is capable of representing a linear combination of $2^n$ basis states, while the classical register requires $n$-bits to store only one of the $2^n$ states.

\paragraph{Quantum circuits} Figure~\ref{fig:framework} (\textit{bottom}) presents an example QNN quantum circuit for QML. It first encodes the raw input through rotation gates like $\mathcal{R}_{\mathrm{X}}$, $\mathcal{R}_{\mathrm{Y}}$, and $\mathcal{R}_{\mathrm{Z}}$. To further manipulate the information in quantum systems, we execute several sequences of (trainable) quantum gates on qubits for desired computations. Specifically, the quantum gates convert one quantum state to another via unitary transformations, i.e., $|\psi\rangle\to\mathcal{U}|\psi\rangle$, where $\mathcal{U}$ is a unitary matrix and can also be parameterized by $\theta$ for adaptive optimizations. Lastly, a qubit readout operation, i.e., measurement, is applied to obtain the prediction from quantum circuits. It collapses a qubit $|\psi\rangle$ from its superposition to either $|0\rangle$ or $|1\rangle$ in probability based on their amplitudes $\alpha$ or $\beta$, respectively. Given a task, designing the best, realistic, noise-robust quantum circuits is highly non-trivial~\cite{ding2020quantum,wang2021quantumnas}, since it needs to express the required transformation (\textit{accurate}), complete the whole execution within an efficient number of steps (\textit{coherent time}), and achieve satisfactory fidelity on the real quantum hardware (\textit{noise adaptive}). 

\paragraph{Variational quantum circuits} If the quantum gates are parameterized by trainable variables, i.e., angles in the quantum rotation gate, this kind of quantum circuit is named the variational circuit. Such parameterized quantum circuits $\Phi(x,\theta)$ usually serve as one of the efficient and robust implementations for diverse quantum applications, which is also our focus in this paper. $\Phi(x,\theta)$ can be adopted to prepare a variational quantum state (i.e., ansatz) as $|\psi(x,\theta)\rangle=\Phi(x,\theta)|0,\cdots,0\rangle$, where $x$ is the input samples of certain quantum application and $\theta$ is the trainable variable. It has demonstrated a great potential in QML~\cite{wittek2014quantum, biamonte2017quantum,schuld2018supervised,benedetti2019parameterized, liang2022variational}, numerical analysis~\cite{lloyd2014quantum, lloyd2016quantum}, quantum simulation \cite{peruzzo2014variational, kandala2017hardware, mcclean2016theory}, and optimizations\cite{moll2018quantum,farhi2014quantum}.

The trainability of variational quantum circuits is one of the key bottlenecks. As presented in~\cite{mcclean2018barren}, a variational circuit might be untrainable with gradient-based approaches due to the excessive flat loss surface (i.e., ``barren plateau''). Meanwhile, the training quality of these circuits directly determines whether the variational quantum algorithm is successful or not. A classic training pipeline of variational quantum circuits contains two parts: ($i$) selecting a handcrafted or searched circuit structure given a computational quantum application; ($ii$) performing the hybrid quantum-classical optimization to update an optimal set of circuit parameters.

The parameterized quantum circuit (PQC) is a popular application of variational quantum circuits, which is broadly used for VQE and QML tasks. Specifically, for VQE tasks, PQC is adopted for preparing states and is measured based on the given molecule. The state preparation and measurement will be executed multiple times on different qubits and bases. An expectation value is calculated as the final result for the estimation of the molecule's ground state energy, as shown in Figure~\ref{fig:vqe_qnn}. As for QML tasks like image recognition, it first encodes the raw pixel with rotation gates (e.g., $\mathcal{R}_{\mathrm{X}}$, $\mathcal{R}_{\mathrm{Y}}$, and $\mathcal{R}_{\mathrm{Z}}$); and then applies parameterized quantum gates to transform the information towards minimum object values; finally, measures the qubits on $\mathrm{Z}$-basis and calculates the according probability for each class with a \texttt{Softmax} function. An illustration of the circuit structure is provided in Figure~\ref{fig:framework}.

To optimize PQCs, the standard way in a simulator is backpropagation, which computes the derivative $\frac{\partial{\mathcal{L}}}{\partial\theta_i}$ of each parameter $\theta_i$ respect to the objective function $\mathcal{L}$, and then update the according parameter as $\theta_{i} = \theta_{i} - \eta\times\frac{\partial \mathcal{L}}{\partial \theta_{i}}$ where $\eta$ is the learning rate. However, this approach is not feasible if we consider the real-noise model or on-chip training due to their non-differentiable nature. To enable the training of quantum circuits in these challenging scenarios, the parameter shift rule is utilized to calculate quantum gradients~\cite{wang2022chip}, which is followed by most of our experiments.

\subsection{QuantumSEA: In-Time Sparse Exploration for Noise-Resistant Quantum Circuits}

\paragraph{QuantumSEA overview} QuantumSEA produces noise-adaptive quantum circuits by conducting in-time sparse explorations. As demonstrated in Figure~\ref{fig:framework}, it contains three different phases: \ding{182} QuantumSEA starts with a randomly sparsified circuit. Each parameter in quantum operations has a probability of being removed or maintained based on certain probability distributions. \ding{183} The sparse circuit is then optimized for several iterations $\Delta \mathrm{T}$, i.e., updating parameters as normal PQC's training. \ding{184} Sparse topologies are changed straight after the weight optimization. Specifically, it first eliminates a portion of quantum gate weights according to the pre-defined pruning criterion and inserts new gates based on the grow indicators. Note that the gate changes from both pruning and growing happen in the logical quantum circuits which will be further compiled before the deployment. 
Lastly, the upgraded sparse quantum circuits are continually trained until the next topology update. A detailed procedure for this in-time exploration is also described in Algorithm~\ref{algo:quantumsea}.


\begin{algorithm}[H]
\caption{In-Time Sparse Exploration for Robust Circuits}
\label{algo:quantumsea}
\renewcommand{\algorithmicensure}{\textbf{Initialize:}}
\begin{algorithmic}[1]
\ENSURE{The quantum circuit $\mathcal{C}(\theta)$, Dataset $\mathcal{D}$, Sparsity distribution $\mathbb{S}=\{s_1,\cdots,s_{\mathrm{B}}\}$, Update schedule \{$\Delta\mathrm{T},\mathrm{T}_{\mathrm{end}},\gamma,f_{\mathrm{decay}}$\}, Learning rate $\eta$}
\STATE{Randomly sparsify circuits with the sparsity $\mathbb{S}$} \COMMENT{\textcolor{gray}{\textit{Initial circuits with reduced quantum gates.}}}
\STATE{\textcolor{gray}{(\textit{Optional})} Initialize gradient magnitude accumulator $\mathcal{M}$}
\FOR{each training iteration $t$}
\STATE Sample a mini-batch from $\mathcal{D}$
\IF{($t$ mod $\Delta\mathrm{T}== 0$) and $t<\mathrm{T}_{\mathrm{end}}$} 
\FOR {each quantum block $b$}
\STATE{$\rho=f_{\mathrm{decay}}(t,\gamma,\mathrm{T}_{\mathrm{end}})\cdot(1-s_b)\cdot \mathrm{N}_b$}
\STATE Prune unnecessary gates and grow new important gates with portion $\rho$ based on certain criteria; Update circuit accordingly  \COMMENT{\textcolor{gray}{\textit{In-time sparse exploration of topologies}}}
\ENDFOR
\ELSE
\STATE Compute gradients through the parameter shift $\nabla_{\theta}\mathcal{L}(\theta)=\frac{1}{2}\times(\mathcal{C}(\theta_{+})-\mathcal{C}(\theta_{-}))\times\frac{\partial\mathcal{L}(\theta)}{\mathcal{C(\theta)}}$~\cite{wang2022chip}
\COMMENT{\textcolor{gray}{\textit{Computing parameter updates under real noise models or on real quantum machines}}}
\STATE{$\theta=\theta-\eta\times\nabla_{\theta}\mathcal{L}(\theta)$} \COMMENT{\textcolor{gray}{\textit{Updating weights}}}
\STATE{\textcolor{gray}{(\textit{Optional})} Update $\mathcal{M}\gets\tau\mathcal{M}+(1-\tau)|\nabla_{\theta}\mathcal{L}(\theta)|$}
\ENDIF
\ENDFOR
\RETURN{A sparse and noise-adaptive quantum circuit} \\
\end{algorithmic}
\end{algorithm}

In our designs, quantum gates are globally ranked according to their significance. Therefore, the sparse topology optimization of different layers is harmonized. Meantime, we do not allow the growth of extra layers, which prevents the increase of circuit depth. Interesting future work is to integrate a circuit depth-based metric during the sparse topology exploration.

The crucial factors of QuantumSEA's performance lie in four aspects: $i$) \textit{Sparsity distribution}, which determines the initial shape of quantum circuits; $ii$) \textit{Update schedule}, that balances the trade-off between exploration and exploitation of sparse circuits' topology; $iii$) \textit{Pruning} and $iv$) \textit{Growth criteria}, which are leveraged to guide the optimization of circuit topologies.

$\rhd$ \textit{Sparsity distribution.} The conventional sparse distribution in sparse training of classical networks, i.e., \textit{Erd$\ddot{o}$s-R$\acute{e}$nyi}~\cite{mocanu2018scalable}, determines the layer-wise sparse ratio to be proportional to the number of parameters within each layer. Directly following this convention, it will degrade to a uniform distribution if each block of quantum circuits has the same number of parameters $\mathrm{N}_b$. However, a na\"ive uniform random sparsification may remove a full control quantum gate, which \textbf{places obstacles to the necessary communications between different qubits} and damages the trainability of resulted circuits. Therefore, we first randomly maintain one parameter for each control gate and then adopt the uniform distribution to sparsify the rest of the parameters. Another potential strategy is to introduce non-parameterized control gates like \texttt{RXYZ} and \texttt{IBMQ Basis} quantum circuits, ensuring essential qubit communication. It is also evidenced by our empirical findings from the sparse topology exploration, the performance of quantum circuits deteriorates when most control gates are pruned, due to the poor entangling capability~\cite{meyer2002global}. 

$\rhd$ \textit{Update schedule.} The update schedule usually contains: 

\begin{itemize}
    \item [($1$)] An update interval $\Delta\mathrm{T}$ that is the number of training iterations between two adjacent updates of sparse topologies.
    \item [($2$)] An ending epoch $\mathrm{T}_{\mathrm{end}}$, after which the sparse connectivity of quantum circuits will be fixed but the updating of their parameters is retained. 
    \item [($3$)] The initial fraction $\gamma$ of quantum gates that allows being pruned or grown. In our case, we set $\gamma=50\%$ via a hyperparameter grid search. It is worth mentioning that QuantumSEA does not introduce thresholds to determine the amount of pruned/grown quantum gates. Instead, it removes or inserts operations with top $\gamma$ insignificance or significance respectively.
    \item [($4$)] A decay schedule $f_{\mathrm{decay}}(t,\gamma,\mathrm{T}_{\mathrm{end}})$ which computes the portion of changeable quantum gate parameters. Following the standard~\cite{evci2020rigging,liu2021we,chen2021chasing}, we select a cosine annealing schedule $\frac{\gamma}{2}\times(1+\mathrm{cos}\frac{t\times\pi}{\mathrm{T}_{\mathrm{end}}})$.
\end{itemize}

$\rhd$ \textit{Pruning criterion.} Several common heuristics for approximating the model parameter's significance can be utilized, including the following three criteria:

\begin{itemize}
    \item [($1$)] Weight magnitude $|\theta|$~\cite{chen2021chasing};
    \item [($2$)] Gradient magnitude $|\nabla_{\theta}\mathcal{L}(\theta)|$~\cite{evci2020rigging,chen2021chasing};
    \item [($3$)] Salience $|\theta\times\nabla_{\theta}\mathcal{L}(\theta)|$~\cite{chen2021chasing}.
\end{itemize}

Usually, criteria (2-3) lead to better performance, which is consistent with our experimental results in Section~\ref{sec:ablation}. A possible reason is that pruning with criteria (2-3) tends to have high gradients in the proceeding iteration of circuit training, leading to a quick reduction of objective values.

$\rhd$ \textit{Growth criterion.} We adopt the superior indicator from literature~\cite{evci2020rigging,liu2021we,chen2021chasing}, i.e., gradient magnitude, to guide the growth of newly added quantum gates. Such criterion normally requires the full model's gradient at each stage for updating topology. However, it is not feasible in quantum circuit training, because increased quantum gates will introduce extra gate noises and may violate the constraint of quantum coherent time. Thus, we are unable to get the gradient information of gates that are not contained in the current sparse circuit, via parameter shift. To tackle this dilemma, we propose a historical gradient accumulator $\mathcal{M}$, which records the gradient of a quantum gate if it has appeared once during the in-time sparse exploration. The accumulator $\mathcal{M}$ is updated by $\tau\mathcal{M}+(1-\tau)|\nabla_{\theta}\mathcal{L}(\theta)|$, where $\tau$ controls the contribution of historical gradients and we use $\tau=0.9$. This design enables a \textit{fixed} and \textit{efficient} number (a small portion) of quantum gates across the overall procedure. Because \ding{182} the size of the accumulator $\mathcal{M}$ is only linearly enlarged along with the increase of quantum qubits, and this negligible memory overhead can be easily afforded by classical computers; \ding{183} The historical gradient information offers an approximated importance of quantum operations, discarding the need of computing the significance for all quantum gates at each training iteration.

Meanwhile, to enhance the exploration ability of QuantumSEA, we further inject randomness into the growth indicator as $\kappa\times r+(1-\kappa)\times \widehat{|\mathcal{M}|}$ (i.e., ``Random + Historical Gradient''). $r$ is a random score drawn from the uniform distribution $\mathcal{U}(0,1)$, $\widehat{|\mathcal{M}|}$ presents the normalized gradient score lied in $[0,1]$, and the coefficient $\kappa$ balances the exploration and exploitation of identifying robust sparse quantum circuits. In our case, $\kappa$ is linearly decayed from $1$ to $0$ along with the training. Note that the number of newly activated quantum gates or parameters is equal to the number of pruned ones, and the new grates are initialized to zero since it produces better results as suggested by~\cite{evci2020rigging,liu2021we,chen2021chasing}.

Note that both pruning and growth are conducted between training iterations when the quantum states have already collapsed. Thus such operations aren't constrained by coherence time. Also, the number of updated gates via pruning and growth is the same. Hence the scale of sparse circuits won't gradually increase and so as the risk of qubit decoherence.

\paragraph{Implicit capacity} We measure the additional implicit circuit capacity $\mathcal{I}$ gained from in-time sparse exploration, which refers to the overall dimensionality of quantum circuits explored in the space-time manifold. To be specific, $\mathcal{I}=$ \{\# the number of gates ever appeared\} / \{\# the number of gates in the sparse circuit\}$-1$ , and $\mathcal{I}$ lies in the range $[0,\frac{1}{1-s}-1]$ where $s\in(0,1]$ is the circuit's sparsity. Our extensive experiments demonstrate that a larger $\mathcal{I}$ empowers circuits with improved expressiveness and robustness towards quantum noises. Additionally, the key to implicit capacity is dynamic exploration and this metric can naturally be generalized beyond QuantumSEA.

\paragraph{Circuits' sparsity} A quantum circuit with higher sparsity is less affected by quantum noise as it has a smaller number of gates and thus less inserting noise on real QC devices, which makes it more trainable in terms of noise tolerance. However, fewer gates limit the expressiveness of circuits in the meantime, which cause the circuits to be less trainable. Fortunately, in our QuantumSEA, we can maintain the circuits with a small number of gates but grant the circuits considerable implicit capacity. Thus it reaches a better trade-off between expressiveness and noise tolerance.

\section{Experiment}

\subsection{Implementation Details} \label{sec:implementation}

\paragraph{Benchmarks} We experiment on $5$ QML and $2$ VQE common benchmarks. QML tasks are the image recognition including MNIST~\cite{lecun1998gradient} $2$-class (\texttt{3}, \texttt{6}), $4$-class (\texttt{0}, \texttt{1}, \texttt{2}, \texttt{3}); Fashion-MNIST~\cite{xiao2017fashion} $2$-class (\texttt{dress}, \texttt{shirt}), $4$-class (\texttt{t-shirt/top}, \texttt{trouser}, \texttt{pullover}, \texttt{dress}), and Vowel~\cite{deterding1990speaker} $4$-class (\texttt{hid}, \texttt{hId}, \texttt{had}, \texttt{hOd}). All these tasks require $4$ logical qubits. Due to limited real quantum computing resources, we report the test accuracy on $300$ randomly selected test examples, following the conventions in~\cite{wang2021quantumnas}. A center crop is applied to trim input images from size $28\times28$ to $24\times24$, and the averaging pooling is utilized to further downsample them to $4\times4$ for $2$-class and $4$-class recognition. As for Vowel-$4$ dataset, it contains $990$ samples, of which $60\%$ data are used for training, $10\%$ for validation, and $30\%$ for testing, following~\cite{wang2021quantumnas}. We compress the original vowel features via principal component analysis (PCA) and only keep the top-$10$ significant dimensions. 

To convert the classical images and vowel features into their quantum representation, we flatten and encode them with rotation gates, where $4$ logical qubits are used for $2$- and $4$-class image recognition, respectively. Specifically, for $4\times4$ inputs, we use $4$ qubits and $4$ layers with $4$ $\mathcal{R}_{\mathrm{X}}$, $4$ $\mathcal{R}_{\mathrm{Y}}$, $4$ $\mathcal{R}_{\mathrm{Z}}$, and $4$ $\mathcal{R}_{\mathrm{X}}$ quantum gates on each layer. In this way, the $16$ classical values are encoded by the $16$ gates as their rotation phases. Similarly, we encode the $10$ vowel features with $4$ qubits and $3$ layers, using $4$ $\mathcal{R}_{\mathrm{X}}$, $4$ $\mathcal{R}_{\mathrm{Y}}$, and $2$ $\mathcal{R}_{\mathrm{Z}}$ for each layer. Note that different input encoding schemes, \textit{e.g.,} qubit and amplitude embedding, can affect the final performance. For a fair comparison, we follow the same encoding method in~\cite{wang2021quantumnas}. During the measurement stage, we compute the expectation values on the Pauli-Z basis, obtain a value in [$-1,1$] for each qubit, and further process them with \texttt{Softmax} to get probabilities. Particularly for the $2$-class image recognition, we sum the qubits \{$0$ and $1$\}, \{$2$ and $3$\} to obtain two values before \texttt{Softmax}. 

For VQE, the objective is to find the low-energy eigenvalue of a target molecule, which calculates the expectation of the molecule's Hamiltonian by a repetitive measurement. The investigated molecules in this paper are \texttt{H$_2$} and \texttt{H$_2$O}, which are encoded with $2$ and $6$ logical qubits. Their Hamiltonians are transformed from the fermionic form to the qubit form by Bravyi-Kitaev mapping~\cite{bravyi2002fermionic}. 

\begin{figure*}[t]
    \centering
    \includegraphics[width=1\linewidth]{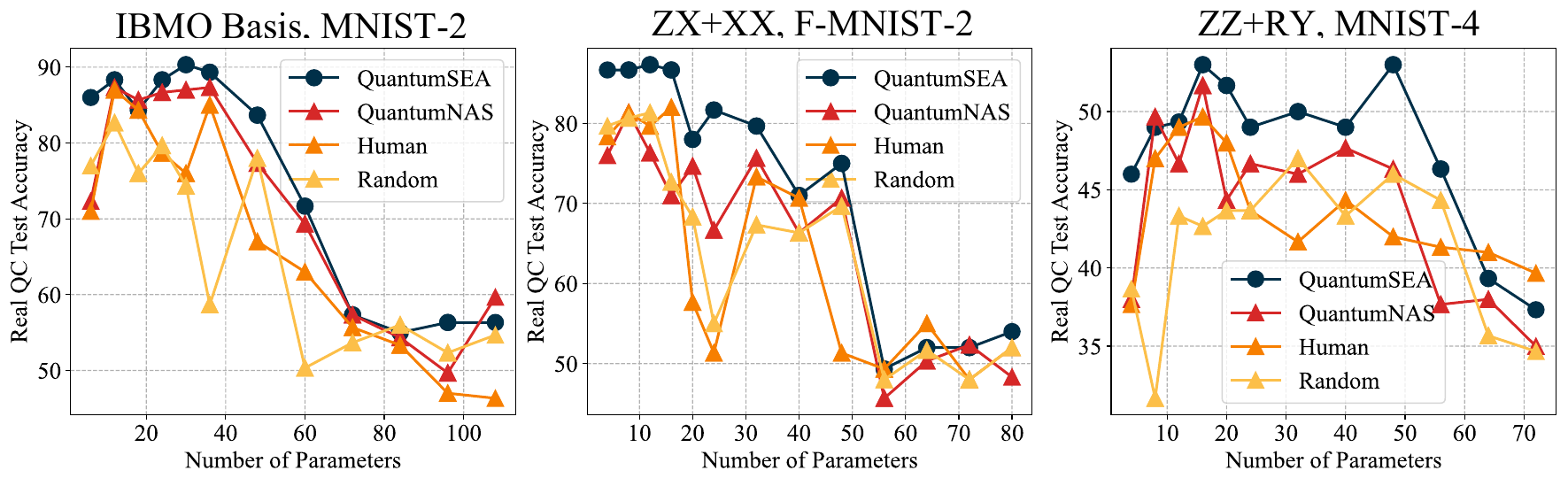}
    \vspace{-3mm}
    \caption{Real QC testing accuracy (\%) over the number of parameters in quantum circuits. Evaluation are conducted on \texttt{IBMQ-Santiago} with configurations of \{(\texttt{IBMQ Basis}, MNIST-2), (\texttt{ZX+XX}, Fashion-MNIST-2), (\texttt{ZZ+RY}, MNIST-4)\} and real noise models~\cite{wang2021quantumnas}.} 
    \label{fig:acc_para_setting}
    \vspace{-1mm}
\end{figure*}

\paragraph{Baseline circuit architectures} \textbf{We inherit the representative circuit designs from~\cite{wang2021quantumnas}}. $i)$ \texttt{U3+CU3} - Each block contains a \texttt{U3} and a \texttt{CU3} layer. The former has one \texttt{U3} gate for each qubit; the latter is constructed with ring connections of \texttt{CU3($0,1$)}, \texttt{CU3($1,2$)}, \texttt{CU3($2,3$)}, and \texttt{CU3($3,0$)}. $ii)$ \texttt{ZZ+RY} - Each block has one \texttt{ZZ} layer with ring connections and one \texttt{RY} layer. $iii)$ \texttt{RXYZ} - Each block contains one \texttt{RX}, \texttt{RY}, \texttt{RZ} and \texttt{CZ} layer, in which the \texttt{CZ} layer is formed with ring connections. Besides, there is one $\sqrt{\texttt{H}}$ layer before the blocks. $iv)$ \texttt{ZX+XX} - Each block contains one \texttt{ZX} and \texttt{XX} layer. $v)$ \texttt{IBMQ Basis} - Each block has \texttt{RZ}, \texttt{X}, \texttt{RZ}, \texttt{SX}, \texttt{RZ}, and \texttt{CNOT} $6$ layers in order. For circuits $i\sim iv)$, we by default use $8$ blocks, except $20$ blocks for circuit $v)$. And we follow the same human design architecture as in~\cite{wang2021quantumnas}.

\paragraph{Quantum noise and associated training configurations} We examine four different quantum noise levels from the ideal to realistic cases. 

\begin{itemize}
    \item [\ding{182}] \textit{Noise-free Simulation}. It does not involve any noise information in training. 
    \item [\ding{183}] \textit{Differentiable Noise Simulation}. Training circuits with approximated noise information~\cite{wang2021roqnn}, which includes quantum error gate insertion via Pauli Twirling~\cite{wang2021roqnn} and readout noise injection.
    \item [\ding{184}] \textit{Real Noise Models}. It adopts the practical noise models from real quantum computers during training.
    \item [\ding{185}] \textit{On-Chip Training}. Training quantum circuits on real quantum devices. 
\end{itemize}

For \ding{182} \textit{Noise-free} and \ding{183} \textit{Differentiable Noise Simulations}, we use an Adam optimizer with $1\times10^{-4}$ weight decay and train the quantum circuits via back-propagation on full training datasets. The learning rate starts from $0$ and linearly increases to $5\times 10^{-3}$ in the first $30$ epochs for QML tasks and $150$ steps for VQE tasks, then decreases by cosine annealing. In total, we train $200$ epochs with batch size $256$ for QML and $1,000$ steps with batch size $1$ for VQE tasks. In QuantumSEA, we update the circuit topology every $10$ epoch for QML and $50$ steps for VQE. For \ding{184} \textit{Real Noise Models} and \ding{185} \textit{On-Chip Training}, we adopt parameter shift~\cite{crooks2019gradients,wang2022chip} to optimize quantum circuits. The learning rate starts from $0.3$ and decays to $0.03$, following a cosine annealing scheduler. We train $50$ epochs and update the structure of circuits every $5$ epoch for QML, while training $250$ steps and updating every $25$ step for VQE. In all settings, we perform pruning and growth only during the front $70\%$ training period, leaving enough training budget for the final sparse circuit to converge. Following~\cite{wang2022chip}, MNIST and Fashion $2$-class adopt the front $500$ images for training; MNIST and Fashion $4$-class choose the front $100$ images as the training set. It is worth mentioning that most of our experiments ($>90\%$) are conducted on practical, and challenging noise-adaptive scenarios and all evaluations are implemented on \textit{real} quantum devices. 

\paragraph{Quantum devices and compiler configurations} TorchQuantum library~\cite{wang2021quantumnas} is used for circuit construction and training. IBMQ quantum computers with Qiskit\footnote{\url{https://quantum-computing.ibm.com/}} APIs are adopted in experiments.


\begin{figure}[t]
    \centering
    \vspace{-2mm}
    \includegraphics[width=0.70\linewidth]{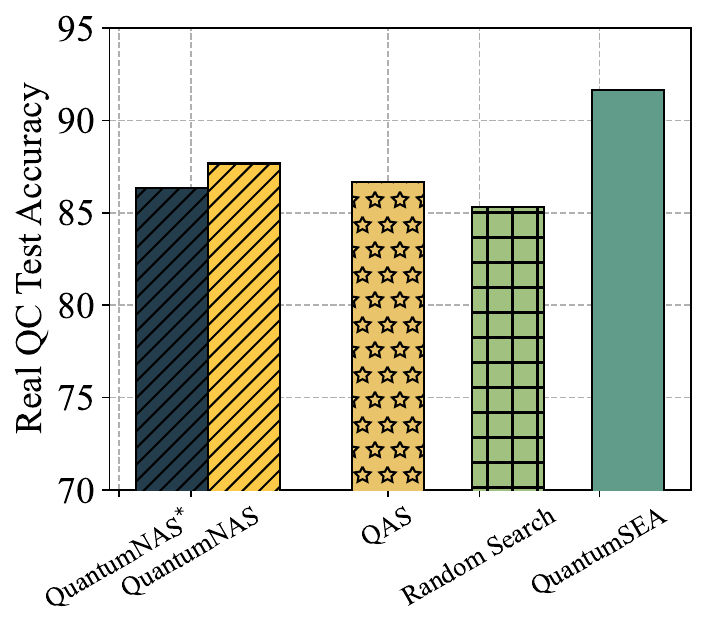}
    \vspace{-3mm}
    \caption{\small QuantumSEA v.s. Search Methods like QuantumNAS~\cite{wang2021quantumnas}, QAS~\cite{du2022quantum}, and Random Search.} 
    \label{fig:beat_nas}
    \vspace{-1mm}
\end{figure}

\subsection{In-Time Sparse Exploration Enables Robust Quantum Circuits}

\textbf{\textit{Q1: What are the key benefits of QuantumSEA? Improved (implicit) capacity and topology of circuits.}}

\begin{table*}[t]
\centering
\caption{Real QC testing accuracy (\%) and energy across diverse quantum tasks (QML and VQE) and circuit ($8$ blocks) types. \textit{Real Noise Simulation} uses real noise models~\cite{wang2021quantumnas} and the parameter-shift update rule during training. \colorbox{lightgray}{\textit{On-Chip Training}} optimizes quantum circuits with the parameter-shift update rule on real quantum devices (the \textit{bottom right} table). \texttt{N.A.} stands for ``not applicable''. A larger accuracy or a smaller energy value implies better performance.}
\vspace{-2mm}
\resizebox{\linewidth}{!}{
\begin{tabular}{@{}l|l|ccccccc}
\toprule
Circuit & \multicolumn{1}{l|}{Method (Sparsity)}  & MNIST-$2$ & MNIST-$4$ & F-MNIST-2 & F-MNIST-4 & Vowel-4 & \texttt{H$_2$} & \texttt{H$_2$O} \\ \midrule
\multirow{10}{*}{\begin{tabular}[l]{@{}l@{}}\texttt{RXYZ} on \\ \texttt{IBMQ-Quito} \end{tabular}} & Human Design ($0\%$) & $88.67$ & $47.67$ & $84.33$ & $64.00$ & $28.45$ & $-1.7263$ & $-36.2711$ \\
& Human Design ($50\%$) & $82.00$ & $55.00$ & $81.67$ & $44.33$ & $31.03$ & $-1.7275$ & $-33.7164$ \\
& Random ($50\%$) & $89.00$ & $56.67$ & $84.00$ & $65.00$ & $44.83$ & $-1.7285$ & $-39.4637$\\
& QuantumNAS ($50\%$)~\cite{wang2021quantumnas} & $88.33$ & $56.33$ & $85.33$ & $66.67$ & $40.52$ & - & -\\ \cmidrule{2-9}
& \textbf{QuantumSEA} ($50\%$) & $\mathbf{90.67}$ & $\mathbf{57.67}$ & $\mathbf{86.00}$ & $\mathbf{72.33}$ & $\mathbf{50.86}$ & $\mathbf{-1.7444}$ & $\mathbf{-39.8873}$\\ 
\cmidrule{2-9}
& Human Design ($0\%$) & $88.67$ & $47.67$ & $84.33$ & $\mathbf{64.00}$ & $28.45$ & $-1.7263$ & $-36.2711$ \\
& Human Design ($80\%$) & $82.67$ & $45.33$ & $75.33$ & $50.00$ & $43.10$ & $-1.7710$ & $-36.4497$ \\
& Random ($80\%$) & $86.67$ & $21.00$ & $88.00$ & $58.00$ & $40.52$ & $-1.6298$ & $-24.4218$ \\
& QuantumNAS ($80\%$)~\cite{wang2021quantumnas}  & $86.00$ & $46.67$ & $83.00$ & $51.67$ & $31.03$ & $-1.3362$ & $-31.3603$ \\ \cmidrule{2-9}
& \textbf{QuantumSEA} ($80\%$) & $\mathbf{89.33}$ & $\mathbf{54.00}$ & $\mathbf{89.33}$ & $62.67$ & $\mathbf{49.14}$ & $\mathbf{-1.7760}$ & $\mathbf{-40.8411}$ \\ 
\midrule
Task & \multicolumn{1}{l|}{Method (Sparsity)} & \texttt{ZZ+RY} & \texttt{RXYZ} & \texttt{ZX+XX} & \texttt{IBMQ Basis} & \multicolumn{1}{|c|}{\colorbox{lightgray}{Circuit}} & \colorbox{lightgray}{MNIST-2} & \colorbox{lightgray}{MNIST-4} \\ \midrule
\multirow{5}{*}{\begin{tabular}[l]{@{}l@{}}MNIST-4 on \\ \texttt{IBMQ-Santiago} \end{tabular}} & Human Design ($0\%$) & $27.67$ & $60.00$ & $34.33$ & $46.00$ & \multicolumn{1}{|c|}{\multirow{5}{*}{\colorbox{lightgray}{\texttt{RXYZ}}}} & $85.00$ & $58.00$\\
& Human Design ($50\%$) & $46.67$ & $52.67$ & $22.33$ & $30.33$ & \multicolumn{1}{|c|}{} & $81.00$ & $53.33$\\
& Random ($50\%$) & $44.67$ & $50.33$ & $42.00$ & $33.00$ & \multicolumn{1}{|c|}{} & $78.67$ & $54.67$ \\
& QuantumNAS ($50\%$)~\cite{wang2021quantumnas} & $\mathbf{51.67}$ & $58.67$ & $44.67$ & $40.00$ & \multicolumn{1}{|c|}{} & \texttt{N.A.} & \texttt{N.A.} \\ \cmidrule{2-6} \cmidrule{8-9}
& \textbf{QuantumSEA} ($50\%$) & $50.33$ & $\mathbf{62.33}$ & $\mathbf{45.67}$ & $\mathbf{60.67}$ & \multicolumn{1}{|c|}{} & $\mathbf{86.33}$ & $\mathbf{55.33}$ \\
\bottomrule
\end{tabular}}
\label{tab:across_qc_tasks}%
\vspace{-1mm}
\end{table*}%

\begin{table}[t]
\centering
\caption{Real QC testing accuracy (\%) across two 6-class classification tasks on \texttt{IBM-OSLO} with \texttt{ZZ+RY} circuits.}
\vspace{-2mm}
\resizebox{\linewidth}{!}{
\begin{tabular}{@{}l|l|cc}
\toprule
Circuit & \multicolumn{1}{l|}{Method (Sparsity)} & MNIST-6 & F-MNIST-6 \\ \midrule
\multirow{5}{*}{\begin{tabular}[l]{@{}l@{}}\texttt{ZZ+RY} on \\ \texttt{IBM-OSLO} \end{tabular}} & Human Design ($0\%$) & $44.67$ & $50.67$  \\
& Human Design ($50\%$) & $39.00$ & $49.00$ \\
& Random ($50\%$) & $43.67$ & $51.00$ \\
& QuantumNAS ($50\%$)~\cite{wang2021quantumnas} & $44.33$ & $52.67$ \\
\cmidrule{2-4}
& \textbf{QuantumSEA} ($50\%$) & $\mathbf{45.33}$ & $\mathbf{54.33}$ \\ 
\bottomrule
\end{tabular}}
\vspace{-4mm}
\label{tab:class_6}
\end{table}

To figure out the key benefits of QuantumSEA, we evaluate it on several QML tasks with \texttt{ZZ+RY} and \texttt{IBMQ Basis} quantum circuits under practical noise models of \texttt{IBMQ-Santiago}. As shown in Figure~\ref{fig:acc_para_setting}, we compare our proposal to circuits with various capacities, obtained from noise-aware searching (QuantumNAS~\cite{wang2021quantumnas}), random generation, and human design. Based on the results, several consistent observations can be drawn:

\begin{itemize}
    \item [\ding{182}] \textit{QuantumSEA delays the accuracy peak.} Along with the rise of circuit parameters, measured real QC testing accuracy first increases to a peak value and then degrades. It is within the expectation of the performance improvements in the beginning, thanks to the enlarged circuit capacity. Notably, excessive parameters of further added quantum gates amplify the noises, leading to deteriorated circuits. However, QuantumSEA maintains fewer gates over the whole training, and leverages dynamic topology updates to explore high-quality sparse quantum circuits. In this way, it expands the in-time circuit capacity implicitly while avoiding additional quantum noises, which allows higher accuracies and postponed performance saturation. 
    \item [\ding{183}] \textit{Comparison with existing approaches.} As indicated by dark curves in Figure~\ref{fig:acc_para_setting}, QuantumSEA consistently surpasses the other three baselines at all parameter levels for diverse scenarios. For example, up to \{$30.66\%$, $30.34\%$, $17.33\%$\} accuracy boosts given the same number of parameters and \{$3.00\%\sim 7.61\%$, $5.33\%\sim 6.00\%$, $1.33\%\sim 6.00\%$\} improvements for the best achievable accuracy on the settings of \{(\texttt{IBMQ Basis}, MNIST-2), (\texttt{ZX+XX}, Fashion-MNIST-2), (\texttt{ZZ+RY}, MNIST-4)\}. It suggests the superiority of located circuit topology against noises, compared to the searched, randomly generated, and human-designed architectures. Therefore, we conclude that the improved implicit capacity and topology of circuits make major contributions.
\end{itemize}

\textbf{\textit{Q2: Does QuantumSEA perform better than neural architecture search-based methods?}}

\begin{table*}[!ht]
\centering
\caption{Real QC testing accuracy (\%) and energy across diverse quantum computing devices with the circuit \texttt{RXYZ}. Real noise models and the parameter-shift update are adopted during training.}
\vspace{-2mm}
\resizebox{\linewidth}{!}{
\begin{tabular}{@{}l|l|c|cccccc}
\toprule
\multicolumn{2}{l|}{Quantum Computing} & Sparsity  & \texttt{IBMQ-Santiago} & \texttt{IBMQ-Bogota} & \texttt{IBMQ-Lima} & \texttt{IBMQ-Belem} & \texttt{IBMQ-Quito} & \texttt{IBMQ-Manila} \\ \midrule
\multirow{5}{*}{\scalebox{0.7}{MNIST-4}} & Human Design & $0\%$ & $60.00$ & $\mathbf{65.00}$ & $54.00$ & $23.67$ & $47.67$ & $68.33$ \\
& Human Design & $\mathbf{50}\%$ & $52.67$ & $53.00$ & $46.33$ & $23.67$ & $55.00$ & $67.67$ \\
& Random & $\mathbf{50}\%$ & $50.33$ & $59.67$ & $56.67$ & $42.00$ & $56.67$ & $58.67$ \\
& QuantumNAS~\cite{wang2021quantumnas} & $\mathbf{50}\%$ & $58.67$ & $57.67$ & $55.67$ & $51.33$ & $56.33$ & $61.00$ \\ \cmidrule{2-9}
& \textbf{QuantumSEA} & $\mathbf{50}\%$ &  $\mathbf{62.33}$ & $62.67$ & $\mathbf{58.67}$ & $\mathbf{52.67}$ & $\mathbf{57.67}$ & $\mathbf{69.67}$\\ \midrule
\multirow{5}{*}{\texttt{H$_2$}} & Human Design & $0\%$ & $-1.6823$ & $-1.6520$ & $-1.7600$ & $-1.4168$ & $-1.7263$ & $-1.7202$  \\
& Human Design & $\mathbf{50}\%$ & $-1.6673$ & $-1.6792$ & $-1.7561$ & $-1.4535$ & $-1.7275$ & $-1.7097$ \\
& Random & $\mathbf{50}\%$ & $-1.6742$ & $-1.6585$ & $-1.7357$ & $-1.4356$ & $-1.7285$ & $-1.7482$ \\
\cmidrule{2-9}
& \textbf{QuantumSEA} & $\mathbf{50}\%$ & $\mathbf{-1.6822}$ & $\mathbf{-1.6989}$ & $\mathbf{-1.7717}$ & $\mathbf{-1.4919}$ & $\mathbf{-1.7444}$ & $\mathbf{-1.7587}$ \\ 
\bottomrule
\end{tabular}}
\label{tab:across_qc_devices}%
\vspace{-1mm}
\end{table*}%

We study our found sparse circuits versus searched circuits, to reveal the advantages of QuantumSEA. The comparison results of (\texttt{RXYZ}, MNIST-2) with $36$ quantum gates, are collected in Figure~\ref{fig:beat_nas} where $\star$ denotes \textit{noise-free} simulation, and the others are under practical noise models from \texttt{IBMQ-Santiago}. We observe that: 

\begin{itemize}
    \item [\ding{182}] The noise-aware optimization brings improved performance, echoed with the finding in~\cite{wang2021quantumnas};
    \item [\ding{183}] Compared to all other three types of searched quantum circuits, QuantumSEA establishes a superior quantum circuit ($\ge 4.00\%$ higher accuracy) by jointly learning topology and weights, which demonstrates the effectiveness of our proposed framework.
\end{itemize}

Note that QuantumNAS and another neural architecture search-based method QAS~\cite{du2022quantum} require a densely optimized super-circuit, suffering from the scalability issue for on-chip training. In contrast, our method always deals with a sparse circuit. Moreover, we update both weights and topologies under real quantum noise, while QuantumNAS and QAS only perform noise-adaptive topology updates.

\textbf{\textit{Q3: Does QuantumSEA achieve superior performance across quantum tasks and circuit types?}}

Comprehensive experiments about different quantum tasks (QML and VQE) and circuit types have been carried out to support the great generalization ability of QuantumSEA. According to the results in Table~\ref{tab:across_qc_tasks} and Table~\ref{tab:class_6}, we see that: 

\begin{itemize}
    \item [\ding{182}] \textit{Win other baselines with improved in-time capacity $\mathcal{I}$ and circuit topologies.} On $5$ QML tasks, $2$ VQE tasks, and $4$ circuit types, QuantumSEA reaches first-rate performance compared to other circuits with the same parameters (e.g., $50\%$ sparsity), implying the better quality of our explored circuit connectivities. In most cases like \texttt{RXYZ} on F-MNIST-2/4 and \texttt{H$_2$} etc., our sparse circuits outperform their dense counterparts ($0\%$ sparsity) significantly. The main reasons stem from two aspects: $i$) excessive quantum gates in dense human-designed circuits induce overmuch noise and hurt the QNN prediction; $ii$) in these cases, QuantumSEA normally obtains sufficient implicit capacity $\mathcal{I}$ ($\ge 0.95$ and $\ge 3.95$ for $50\%$ and $80\%$ sparsity circuits), signifying a thorough exploration of space-time manifold and enhanced expressiveness.
    \item [\ding{183}] \textit{Superior performance in on-chip training.} Based on a recent state-of-the-art on-chip training research~\cite{wang2022chip}, we examine our approaches on MNIST-2/4 with \texttt{RXYZ} circuits. \colorbox{lightgray}{Marked} results (Table~\ref{tab:across_qc_tasks}) suggest that ours surpass human-designed circuits~\cite{wang2022chip} (both $50\%$ and $100\%$) by a margin of $1.33\%\sim 5.32\%$ performance improvements on MNIST-2, and win $1.99\%$ accuracy gains on MNIST-4 compared to human-designed circuits with the same number of quantum gates.
    \item [\ding{184}] For a more complicated 6-class classification task. QuantumSEA also achieves superior performance compared with other circuits with the same parameters. Specifically, as shown in Table~\ref{tab:class_6}, QuantumSEA obtains an extra performance improvement by up to $6.33\%$ and $5.33\%$ on MNIST-6 and F-MNIST-6, respectively.
\end{itemize}

\soulregister\ref{7}

Note that for a fair comparison, we adopt the same circuit design style from~\cite{wang2021quantumnas}, which explores the circuit topology within a basic block and duplicates it to construct the full quantum circuits. As shown in Table~\ref{tab:across_qc_tasks}, our methods achieve consistent performance gains in most cases.

\textbf{\textit{Q4: Does QuantumSEA achieve superior performance across diverse real quantum devices?}} 

We further evidence the superiority of QuantumSEA across six different QC machines. For each quantum device, we train circuits with their associated practical noise model and evaluate optimized circuits on real QC. Table~\ref{tab:across_qc_devices} tells that: 

\begin{itemize}
    \item [\ding{182}] Given the same number of quantum gates (or sparsity), our proposals show clear advantages in accuracy gains and energy decrements, indicated by \textbf{bold} numbers.
    \item [\ding{183}] Although the dense ($0\%$ sparsity) human-designed circuits moderately beat QuantumSEA in some QCs like \texttt{IBMQ-Bogota}, our sparse ones still enjoy around $50\%$ time savings for both training and inference, as also depicted in Figure~\ref{fig:acc_para_noise_teaser} (\textit{Right}) and~\ref{fig:coherence_time}. Determined by the physical environments of quantum machines, the operation error of executing quantum gates has varied intensities~\cite{wang2021quantumnas}. It is why certain real QCs might be more amenable to handling circuits with more parameters.
\end{itemize}

\subsection{Ablation Study and Visualization} \label{sec:ablation}

\paragraph{Ablation on pruning and growth criteria} We perform the ablation study on pruning and growth indicators, which play crucial roles in QuantumSEA. From Table~\ref{tab:ablation}, experiments with different sparsity levels show consistent preferences of the criteria. Specifically, using ``Salience'' to estimate weight significance for pruning and adopting the fused score of ``Historical (His.) Gradient + Random'' for growing new connections, demonstrates the best performance, i.e., $1.00\%\sim 12.67\%$ better accuracy compared to other options. Thus, we inherit this optimal setup for all other experiments. Additionally, to evaluate the impact of the historical gradient, we replace it with the vanilla gradient in the optimal setup. We can observe that the historical gradient significantly improves the final accuracy by up to $4.34\%$, which is expectable because using the vanilla gradient, the gates that are not contained in the current circuits can only grow by a random chance.

\begin{table}[t]
\vspace{-1mm}
\caption{\small Ablation of pruning and grow criteria in QuantumSEA. Evaluations are conducted on \texttt{IBMQ-} \texttt{Santiago} with (\texttt{ZZ+RY}, MNIST-4) and practical noise models. ``Ours'' denotes ``His. Gradient + Random''.}
\vspace{-1mm}
\label{tab:ablation}
\centering
\resizebox{0.49\textwidth}{!}{
\begin{tabular}{l|cc|c}
\toprule
\multirow{1}{*}{Sparsity} & \multirow{1}{*}{Pruning} & \multirow{1}{*}{Grow} & \multirow{1}{*}{Accuracy (\%)}  \\  
\midrule
$50$\% & Random & Ours & $43.67$ \\ 
$50$\% & Weight & Ours & $46.67$ \\ 
$50$\% & His. Gradient & Ours & $42.33$ \\
$50$\% & Salience & Ours w.o. His. & $43.33$ \\
\rowcolor[gray]{0.9} 
$50$\% & Salience & Ours & $\mathbf{47.67}$ \\ \midrule
$50$\% & Salience & His. Gradient & $42.33$ \\
$50$\% & Salience & Random & $35.00$ \\ \midrule
$80$\% & Random & Ours & $47.33$  \\ 
$80$\% & Weight & Ours & $49.67$ \\ 
$80$\% & His. Gradient & Ours & $45.00$  \\ 
$80$\% & Salience & Ours w.o. His. & $48.33$ \\
\rowcolor[gray]{0.9} 
$80$\% & Salience & Ours & $\mathbf{51.67}$  \\ \midrule
$80$\% & Salience & His. Gradient & $49.33$ \\
$80$\% & Salience & Random & $45.00$ \\ 
\bottomrule
\end{tabular}}
\vspace{-1mm}
\end{table}

\begin{figure}[t]
\centering
\includegraphics[width=0.9\linewidth]{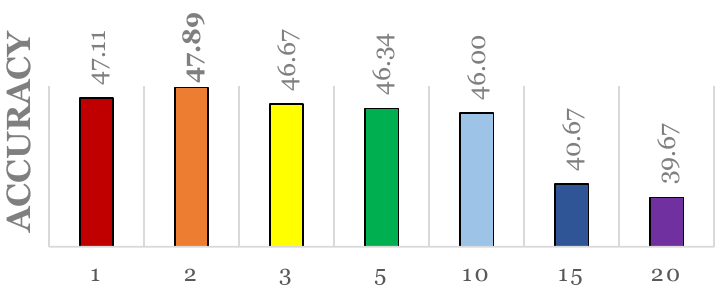}
\vspace{-3mm}
\caption{\small Ablation study of update intervals.} 
\label{fig:ablation_T}
\vspace{-1mm}
\end{figure}

\begin{table}[!ht]
\centering
\caption{\small Compatible with existing techniques of noise mitigation~\cite{wang2021roqnn}. We use the same experiment configurations in \cite{wang2021roqnn}.}
\vspace{-3mm}
\label{tab:other_mitigation}
\resizebox{0.90\linewidth}{!}{
\begin{tabular}{l|cc}
\toprule
Method & F-MNIST-4 (\textbf{Full Training Data}) \\ \midrule
Ours (50\% Sparsity) & $72.33$ \\ \midrule
\qquad + Normalization & $83.00$ \\
\qquad + Quantization  & $74.00$ \\ \midrule
Ours (80\% Sparsity) & $54.67$  \\ \midrule
\qquad + Normalization & $69.00$ \\
\qquad + Quantization  & $66.33$ \\ 
\bottomrule
\end{tabular}}
\end{table}

\begin{figure}[!ht]
    \centering
    \vspace{-2mm}
    \includegraphics[width=0.7\linewidth]{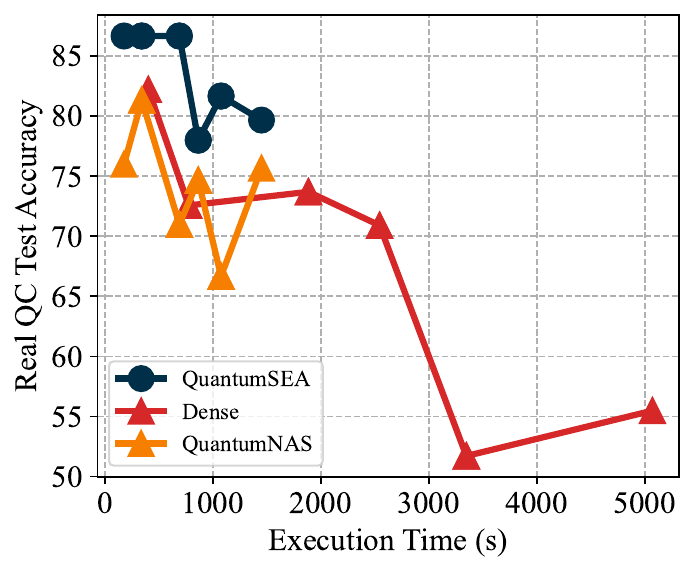}
    \vspace{-3mm}
    \caption{\small Real QC testing accuracy (\%) of Fashion-MNIST-2 over the training time of quantum circuit \texttt{ZX+XX} on \texttt{IBMQ-Santiago}. Our proposed QuantumSEA achieves a better Pareto Frontier approximation of the trade-off between accuracy and execution time.} 
    \label{fig:coherence_time}
\end{figure}

\paragraph{Ablation on the update interval of QuantumSEA} The length of the update interval $\Delta\mathrm{T}$ controls the critical exploration and exploitation trade-off~\cite{liu2021we,chen2021chasing} in QuantumSEA. For instance, a larger $\Delta \mathrm{T}$ emphasizes exploitation since more epochs are tokens to approximate weight significance; on the contrary, a smaller $\Delta \mathrm{T}$ allows more topology updates and enhances the exploration ability (usually comes with a larger implicit capacity $\mathcal{I}$). We conduct the ablation with fine-grained $\Delta \mathrm{T}$ choices in Figure~\ref{fig:ablation_T}. It suggests $\Delta \mathrm{T}=2$ is the ``sweet-point'' setup of our in-time sparse circuit explorations. 

\paragraph{Compatibility with existing noise mitigation} QuantumSEA functions in orthogonal to existing noise mitigation mechanisms, such as post-measurement normalization and quantization from the recent advance~\cite{wang2021roqnn}. The compatibility is shown by obtaining extra performance improvements from their combinations, as recorded in Table~\ref{tab:other_mitigation}.

\paragraph{Discussion on quantum coherence time} Since the runtime on real quantum devices scales nearly linearly to \# qubits and \# quantum gates, QuantumSEA exhibits great advantages in the trade-off between real QC accuracy and execution time, as shown in Figure~\ref{fig:coherence_time}. It is mainly because our methods enable superior performance with a small number of quantum gates with the assistance of in-time implicit capacity.

\paragraph{Ablation on implicit capacity}

\begin{figure}[t]
    \centering
    \includegraphics[width=1\linewidth]{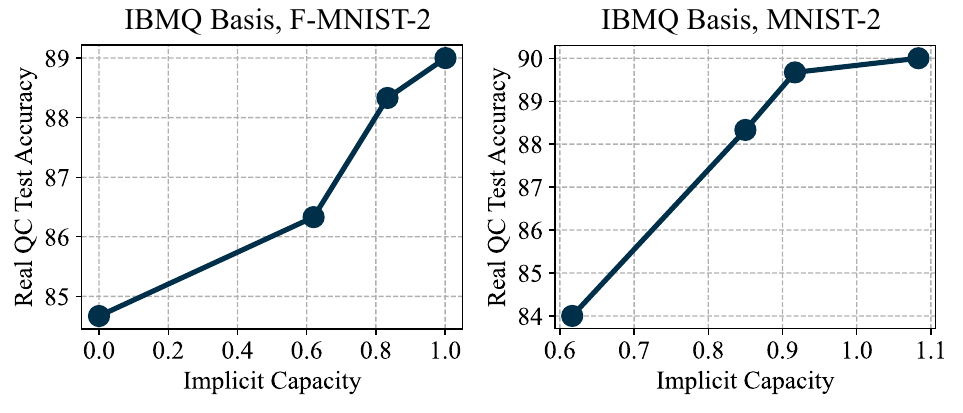}
    \vspace{-3mm}
    \caption{\small Real QC testing accuracy (\%) of Fashion-MNIST-2 and MNIST-2 over the implicit capacity of quantum circuit \texttt{IBMQ Basis} on the \texttt{IBMQ-Quito} quantum machine.}
    \label{fig:implicit}
    \vspace{-1mm}
\end{figure}

Our QuantumSEA provides the quantum circuit training with extra implicit capacity $\mathcal{I}$. As shown in Figure~\ref{fig:implicit}, the granted implicit capacity $\mathcal{I}$ is positively related to the final achievable accuracy. A larger $\mathcal{I}$ promotes circuits with enhanced expressiveness and robustness against quantum errors, which is a potential reason for QuantumSEA's superior performance. On the contrary, as other competitive methods won't dynamically adjust the sparse circuits, their implicit capacity equals zero.

\paragraph{Visualization of dense and sparse quantum circuits} Figure~\ref{fig:circuit_vis} presents identified sparse quantum circuit topologies from QuantumSEA (\textit{Middle}) \& QuantumNAS (\textit{Right}) and their dense counterpart (\textit{Left}). Several observations can be drawn from these results:

\begin{figure}[t]
    \centering
    \includegraphics[width=0.32\linewidth]{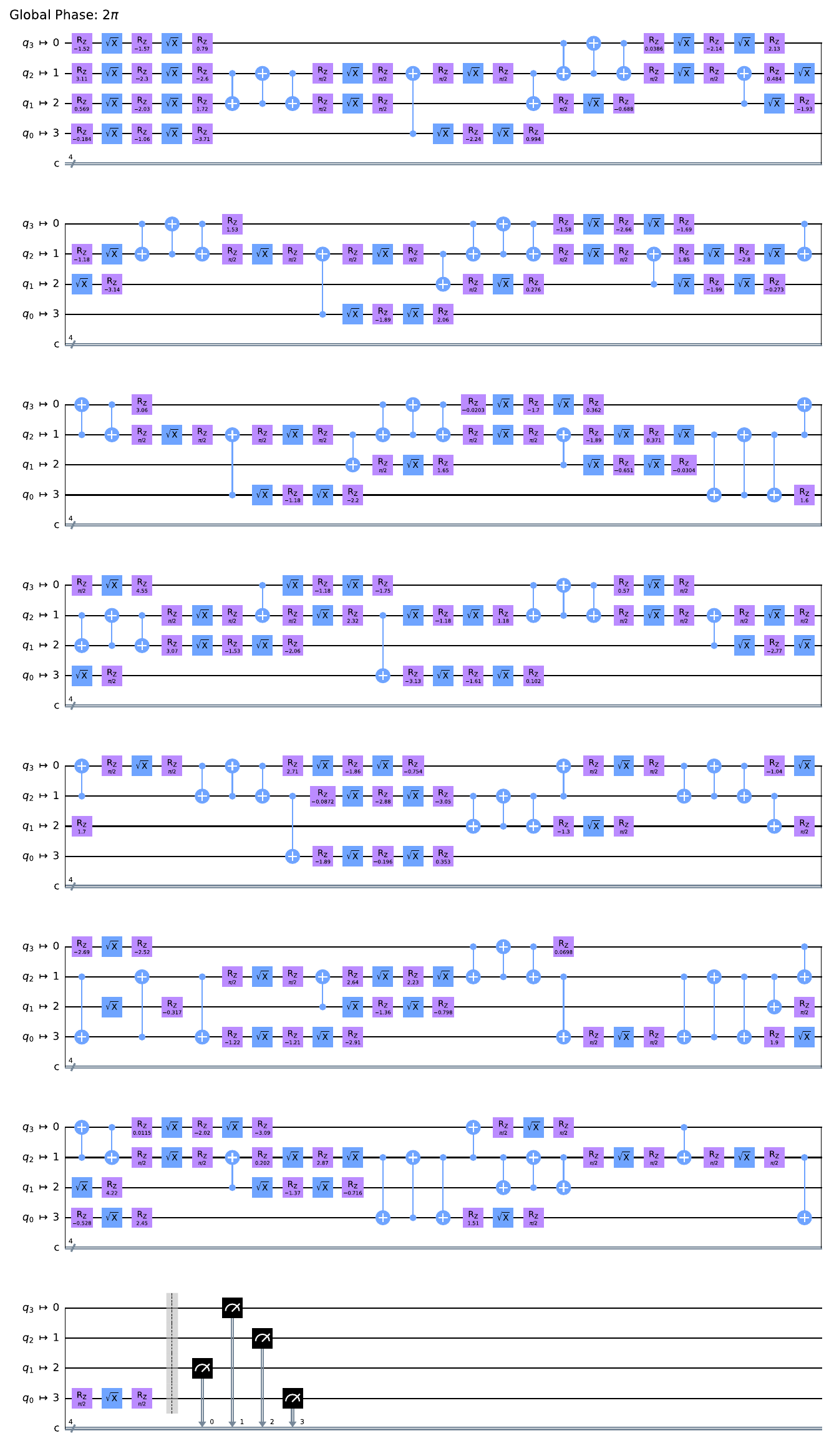}
    \includegraphics[width=0.32\linewidth]{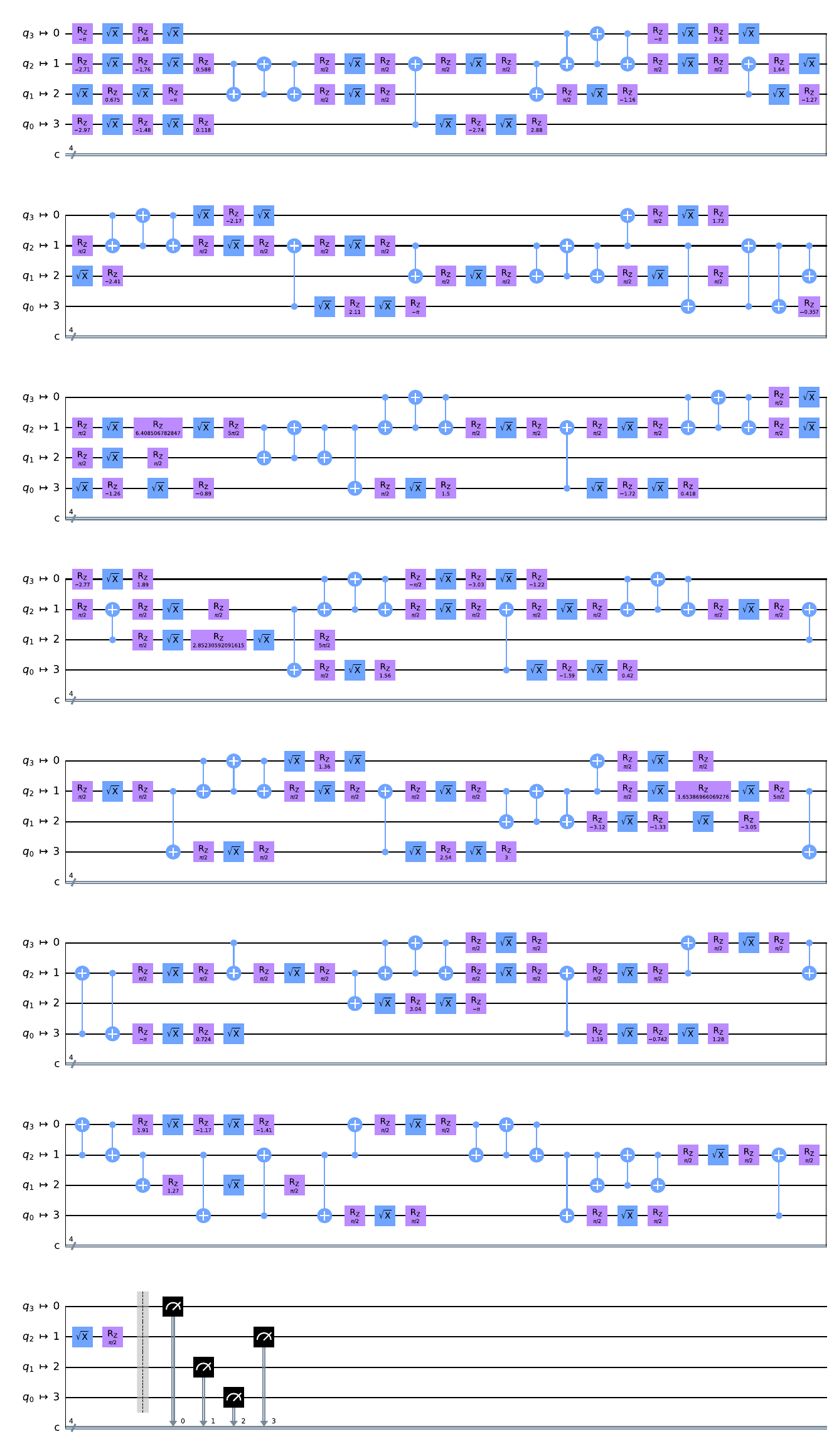}
    \includegraphics[width=0.32\linewidth]{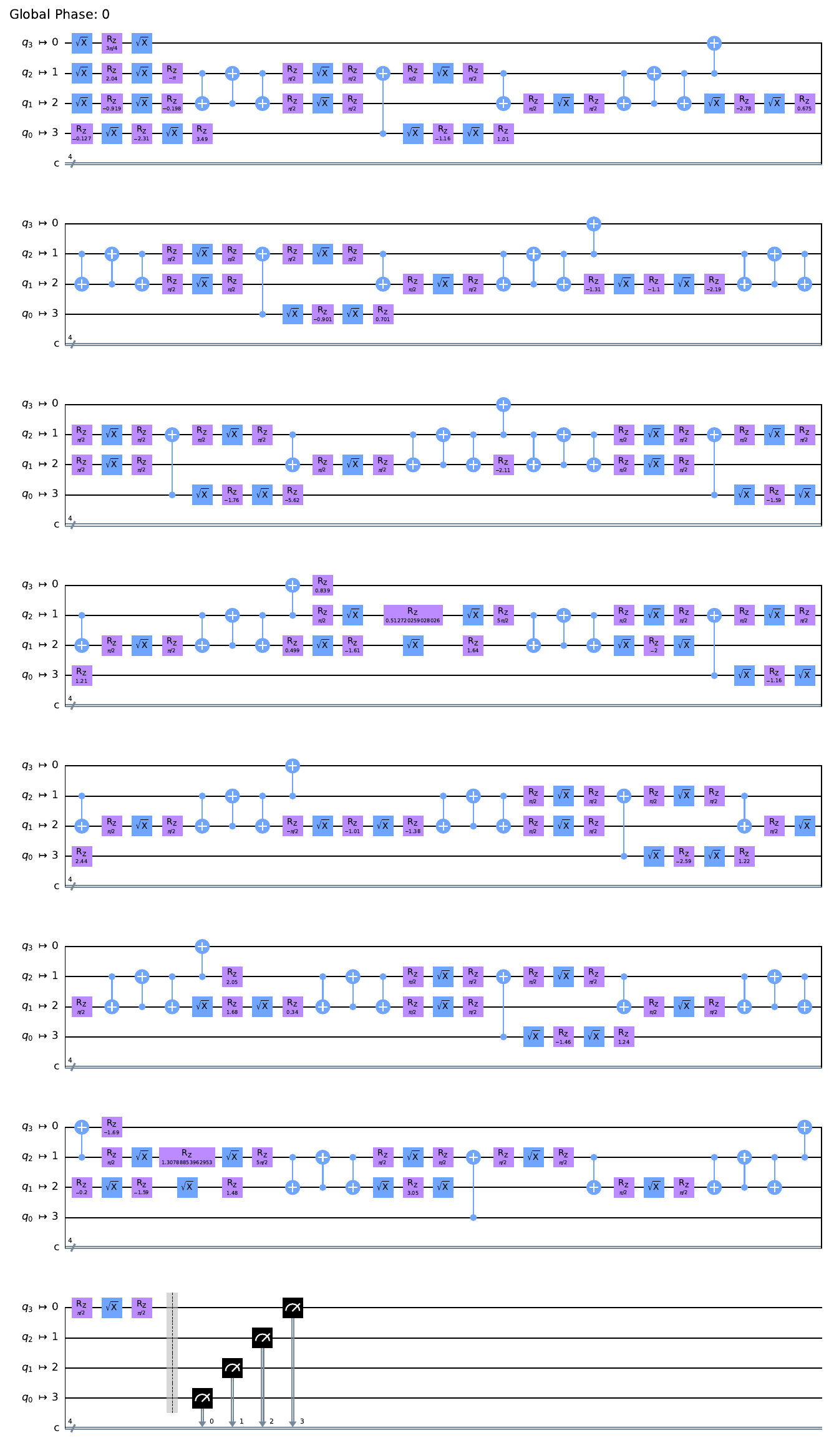}
    \caption{Visualization of a dense circuit (\textit{Left}) and sparse circuits from our QuantumSEA (\textit{Middle}) \& QuantumNAS (\textit{Right}) with $8$ blocks \texttt{RXYZ} on \texttt{IBMQ-Quito}. Please kindly zoom in for better readability.} 
    \vspace{-2mm}
    \label{fig:circuit_vis}
\end{figure}

\begin{itemize}
    \item [\ding{182}] Compared with the dense circuit in Figure~\ref{fig:circuit_vis} (\textit{Left}), the sparse circuit from QuantumSEA (Figure~\ref{fig:circuit_vis} (\textit{Middle})) preserves a similar number of remaining quantum gates for each qubit. In the meantime, most of the control gates across different qubits are maintained. Such ``undamaged" communication ability between qubits could be one possible reason for the superiority of QuantumSEA. 
    \item [\ding{183}] Quantitative results of \{\# depth, \# X/Y/Z gate, \# CZ gate, expressibility~\cite{sim2019expressibility}\} for QuantumNAS and our QuantumSEA are \{$32$, $96$, $24$, $0.0064$\} and \{$31$, $48$, $24$, $0.0047$\}, respectively. It shows the improved expressibility (i.e., smaller Kullback Leibler divergence) and reduced gate error of our generated sparse circuits.
\end{itemize}

\section{Conclusion and Discussion} \label{sec:conclusion}

In this paper, we introduce an in-time sparse exploration framework, QuantumSEA, which jointly optimizes the topology and weight of quantum circuits under practical quantum noises. Its advantages lie in: ($1$) Sparse circuits with reduced quantum gates over the full training and inference processes, substantially alleviating the effects of decoherence; ($2$) The in-time dynamic connectivity exploration produces extra implicit capacity, greatly enhancing the circuit's expressiveness; ($3$) Identified noise-adaptive sparse topology regularizes circuit training and boosts the noise resistance. Comprehensive experiments validate our proposals regarding superior performance and significant training time savings. Our approaches currently only consider trimming quantum gates in the circuit. We plan to investigate the possibility of sparsifying both quantum qubits and gates in future works. Although we see our work as scientific in nature, it might amplify the existing negative societal impacts of using quantum computing because we have no control over anyone who can get access to our QuantumSEA and its generated quantum circuits. 

\textit{When and what tasks do PQCs can achieve quantum advantages?} Currently, quantum approaches have demonstrated advantages in applications like large unitary matrix multiplication. To outperform classical neural networks, we think it requires PQCs with more than $50$ qubits and effective error correction mechanisms (or fault-tolerant machines).

\section*{ACKNOWLEDGMENT}
This work is funded in part by EPiQC, an NSF Expedition
in Computing, under award CCF-1730449; in part
by STAQ under award NSF Phy-1818914; 
in part by the US Department of Energy Office 
of Advanced Scientific Computing Research, Accelerated 
Research for Quantum Computing Program; and in part by the 
NSF Quantum Leap Challenge Institute for Hybrid Quantum Architectures and 
Networks (NSF Award 2016136) and in part based upon work supported by the 
U.S. Department of Energy, Office of Science, National Quantum 
Information Science Research Centers. FTC is Chief Scientist for Quantum Software at Infleqtion and an advisor to Quantum Circuits, Inc. We thank MIT-IBM Watson AI Lab, Qualcomm Innovation Fellowship for supporting this research. We acknowledge the use
of IBM Quantum services for this work.


\bibliographystyle{IEEEtranS}
\bibliography{ref}

\end{document}